\documentclass[prd,twocolumn,balancelastpage]{revtex4}
\usepackage{graphicx,natbib}
\usepackage{hyperref,amsmath}

\begin{document}

%-------------------------------------------------

\title{Evidence for a new light spin-zero boson from cosmological gamma-ray propagation?}

%-------------------------------------------------

\author{Alessandro De Angelis\vspace*{.5mm}}
\altaffiliation{also at IST, Lisboa, Portugal}
\affiliation{Dipartimento di Fisica, Universit\`a di Udine, Via delle Scienze 208, I-33100 Udine, and INAF and INFN, Sezioni di Trieste, Italy}
\author{Marco Roncadelli\vspace*{.5mm}}
\affiliation{Dipartimento di Fisica Nucleare e Teorica, Universit\`a di Pavia and INFN, Sezione di Pavia, Via A. Bassi 6, I-27100 Pavia, Italy}
\author{Oriana Mansutti\vspace*{.5mm}}
\affiliation{Dipartimento di Fisica, Universit\`a di Udine, Via delle Scienze 208, I-33100 Udine, and INFN, Sezione di Trieste, Italy\vspace*{.5mm}\\
{\rm (Received 30 July 2007; revised manuscript received 20 September 2007; published 18 December 2007)}}

%-------------------------------------------------

\begin{abstract}
Recent findings by imaging atmospheric Cherenkov telescopes indicate \mbox{a large transparency of the} Universe to gamma rays, which can be hardly explained within the current models of extragalactic background light. We show that the observed transparency is naturally produced by an oscillation mechanism---which can occur inside intergalactic magnetic fields---whereby \mbox{a photon can become a} new spin-zero boson with mass $m\!\!\ll\!\!10^{-10}\,{\rm eV}$. Because the latter particle travels unimpeded throughout the Universe, photons can reach the observer even if the distance from the source considerably exceeds their mean free path.
We compute the expected flux of gamma rays from blazar 3C279 at different~energies. Our predictions can be tested in the near future by the gamma-ray telescopes H.E.S.S., MAGIC, CANGAROO and VERITAS.
Moreover, our result provides an important observational test for models of dark energy wherein quintessence is coupled to the photon through an effective dimension-five operator.
\\
\\
DOI: 10.1103/PhysRevD.76.121301 \hfill PACS numbers: 95.85.Pw, 14.80.Mz
\end{abstract}

%-------------------------------------------------

\maketitle

%-------------------------------------------------

Thanks to the advent of imaging atmospheric Cherenkov telescopes (IACTs) like H.E.S.S., MAGIC, CANGAROO III and VERITAS, the detection of photons from astrophysical sources in the Very High Energy (VHE) window above a threshold close to 100 GeV and up to some $10\,{\rm TeV}$ has become an exciting reality.
Active galactic nuclei (AGN) are known emitters of photons in that energy range through specific processes arising from their accretion~\cite{fossati}.
A number of such emitters have been observed in the last few years, and about~20~AGN~at~VHE are known today~\cite{persic}.
The redshift $z$ at which thesesources are detected is steadily increasing, and recently the MAGIC collaboration has reported~\cite{3c} an impressive record at~$z=0.538$ with the discovery of the emission from 3C279~\cite{egretdet}.

According to the current understanding, the flux of VHE gamma rays from distant sources is attenuated in an energy-dependent way by the interaction with background photons in the Universe~\cite{stecker1971}. Specifically, the dominant process for the absorption of a photon of 
energy~$E$ is pair-production $\gamma \gamma \to e^+ e^-$. Actually, $\sigma (\gamma \gamma \to 
e^+ e^-)$ becomes maximal for a background photon energy $\epsilon \simeq (500 \, \rm{GeV}/E)\,{\rm eV}$.
In the energy interval $100~{\rm GeV} < E < 1 \, {\rm TeV}$ the absorption is dominated by the interaction with optical/infrared photons of the metagalactic radiation field (MRF), often called extragalactic background light (EBL)~\cite{gould:1967a}, while the interaction with the CMB becomes dominant at \mbox{$E \sim 10^6 \, {\rm GeV}$}.

A general trend seems to emerge from the most recent determinations of the EBL density~\cite{aharonian:nature06,mazin}. Observations of AGN at redshift $z=0.165$ and $z=0.186$ entail that intergalactic space might be more transparent to gamma rays than previously thought.
Since only basic QED and the theory of special relativity are involved in the calculation of 
$\sigma (\gamma \gamma \to e^+ e^-)$, it might be difficult to account for the observed attenuation without invoking new physics beyond the Standard Model~\cite{diffic}.
Even in a model with the lowest EBL density compatible with cosmology, one expects for a source at redshift \mbox{$z \simeq 0.538$} an attenuation by less than 0.5 at 100~GeV, which increases to about \mbox{$e^{-4} \simeq 0.018$} at $E=500$~GeV~\cite{kneiske}. This makes a source like 3C279  hardly visible in the VHE range---or, in any case, causes a severe difference in the attenuation of the signal at, say, 100 GeV, as compared to the attenuation at, say, 500 GeV. Yet, MAGIC has found that the signal from 3C279 collected in the region $E<220$ GeV has more or less the same statistical significance as the signal in the range 220 GeV $< E <$ 600 GeV ($6.1 \sigma$ in the former case, $5.1 \sigma$ in the latter)~\cite{3c}. It seems therefore interesting to explore which kind of new physics would be required to explain the observations of VHE sources.

In this paper, we show that a way out of the above difficulty is provided by a simple mechanism, according to which part of the photons are converted into light spin-zero bosons, which can travel unimpeded through cosmological distances; part of such light bosons are in turn reconverted into photons and detected, so that everything goes as if an anomalously small attenuation were operative. As a result, photons can reach the observer even if their mean free path is considerably smaller than the distance from the source.
We show that this mechanism indeed predicts an effect which is quantitatively adequate to explain the data.

A natural possibility of this kind arises in extensions of the standard model containing a new light spin-zero boson $X$ coupled to the photon through an effective dimension-five operator. In order to be specific, we shall be concerned throughout with the effective Lagrangian
\begin{equation}
\label{aq5}
{\cal L}_{\phi \gamma} = - \frac{1}{4 M} \, F^{\mu \nu} \,  \tilde F_{\mu \nu} \, \phi = \frac{1}{M} \, {\bf E} \cdot {\bf B} \, \phi~,
\end{equation}
where $\phi$ stands for the $X$ field and $M$ is a parameter with the dimension of an energy, playing the role of the inverse of a coupling constant~\cite{SCA}.
A Lagrangian like ${\cal L}_{\phi \gamma}$ appears in a wide class of realistic four-dimensional models~\cite{masso1}, in particular in the phenomenology of axion-like particles, as well as in compactified Kaluza-Klein theories~\cite{kk} and superstring theories~\cite{superstring}.
Moreover, it has been argued that the presence of ${\cal L}_{\phi \gamma}$ should be a generic feature of quintessential models of dark energy~\cite{carroll}. A particular case concerns the axion~\cite{assione}, which is characterized by the mass-coupling relation \mbox{$m \sim ( 10^{10} \, {\rm GeV}/M ) \, {\rm eV}$}. 

Before proceeding further, we recall that astrophysics provides a lower bound on $M$, depending on the mass~$m$ of the $X$~boson.
Failure to detect $X$ bosons from the Sun by the CAST experiment at CERN sets the robust bound \mbox{$M > 1.14 \cdot 10^{10} \, {\rm GeV}$} for \mbox{$m < 0.02 \, {\rm eV}$}~\cite{cast}, which practically coincides with the theoretical bound derived from the properties of globular clusters~\cite{Raffelt1990}.
Furthermore, the stronger bound \mbox{$M > 3 \cdot 10^{11} \, {\rm GeV}$} holds for $X$~bosons  with \mbox{$m < 10^{- 10} \, {\rm eV}$}, based both on the energetics of the supernova 1987a~\cite{raffeltmasso} and on observations of time-lag between opposite-polarization modes in pulsar radio emission~\cite{mohanti}.

Owing to ${\cal L}_{\phi \gamma}$, the interaction eigenstates differ from the propagation eigenstates in the presence of a magnetic field ${\bf B}$, so that ${\gamma}$-$X$ interconversion occurs. Coherent \mbox{${\gamma}$-$X$} mixing can be understood as an oscillation process quite similar to that taking place for massive neutrinos of different flavours, apart from the fact that in the case described in this paper the external~${\bf B}$ field is necessary to account for the spin mismatch~\cite{RaffeltStodolsky}.

As an illustrative example, we compute the fraction of the initial photon flux which survives the distance from the source 3C279 at $z$~=~0.538 when both the absorption from EBL and ${\gamma}$-$X$ oscillations in intergalactic magnetic fields are taken into account~\cite{noi}.

Intergalactic magnetic fields have a complicated and poorly known morphology, which reflects the pattern of baryonic structure formation as well as its subsequent evolutionary history~\cite{magfields}.
However, for the present needs it is sufficient to suppose that ${\bf B}$ is constant over a domain of size $L_{\rm dom}$, with ${\bf B}$ randomly changing its direction from one domain to another but keeping the same strenght. Values to be used throughout are \mbox{$B \simeq 10^{- 9} \, {\rm G}$} and \mbox{$L_{\rm dom} \simeq 1 \, {\rm Mpc}$}, which are close to existing upper limits but consistent with them~\cite{fur}.

Let us consider first the propagation of a photon beam over a single magnetic domain, in the presence of a cold intergalactic plasma with plasma frequency ${\omega}_{\rm pl} = \sqrt{4 \pi \alpha n_e/m_e} \simeq 3.69 \cdot 10^{- 11} \, \sqrt{n_e /{\rm cm}^{- 3}} \, {\rm eV}$, 
where~$n_e$ denotes the electron density. In order to have an unsuppressed amplitude, we work 
in the strong-mixing regime, which requires \mbox{$E \gg |m^2 - {\omega}^2_{\rm pl}|M/2 B$}. A conservative estimate of the density of intergalactic plasma yields \mbox{$n_e \simeq 10^{-7} \, {\rm cm}^{-3}$}~\cite{peebles}, resulting in the plasma frequency \mbox{${\omega}_{\rm pl} \simeq 1.17 \cdot 10^{-14} \, {\rm eV}$}~\cite{wmap}. Therefore, the latter condition takes the explicit form $|(m/10^{- 10} \, {\rm eV})^2 - 1.37 \cdot 10^{- 8}| \ll 0.38 ({\omega}/{\rm GeV}) ( B_T/10^{-9} \, {\rm G})$ $(10^{10} \, {\rm GeV}/M)$.
Since we are interested in the energy range $E > 10^2 \, {\rm GeV}$, we find that for e.g.\ \mbox{$M > 4 \cdot 10^{11} \, {\rm GeV}$} the bound \mbox{$m \ll 10^{- 10} \, {\rm eV}$} has to be satisfied~\cite{noaxion}. We remark that the present mechanism works for arbitrarily small values of $m$, provided $M$ is considerably smaller than the Planck mass $M_P \simeq 1.22 \cdot 10^{19} \, {\rm GeV}$~\cite{planck}. As a consequence, our result also applies to models of dark energy wherein quintessence enjoys a photon coupling described by ${\cal L}_{\phi \gamma}$~\cite{carroll}, thus ultimately providing an important observational test for these models.

\begin{figure}[b!]
\vspace*{-5mm}
\centering \hspace*{-6mm}
\includegraphics[width=.55\textwidth]{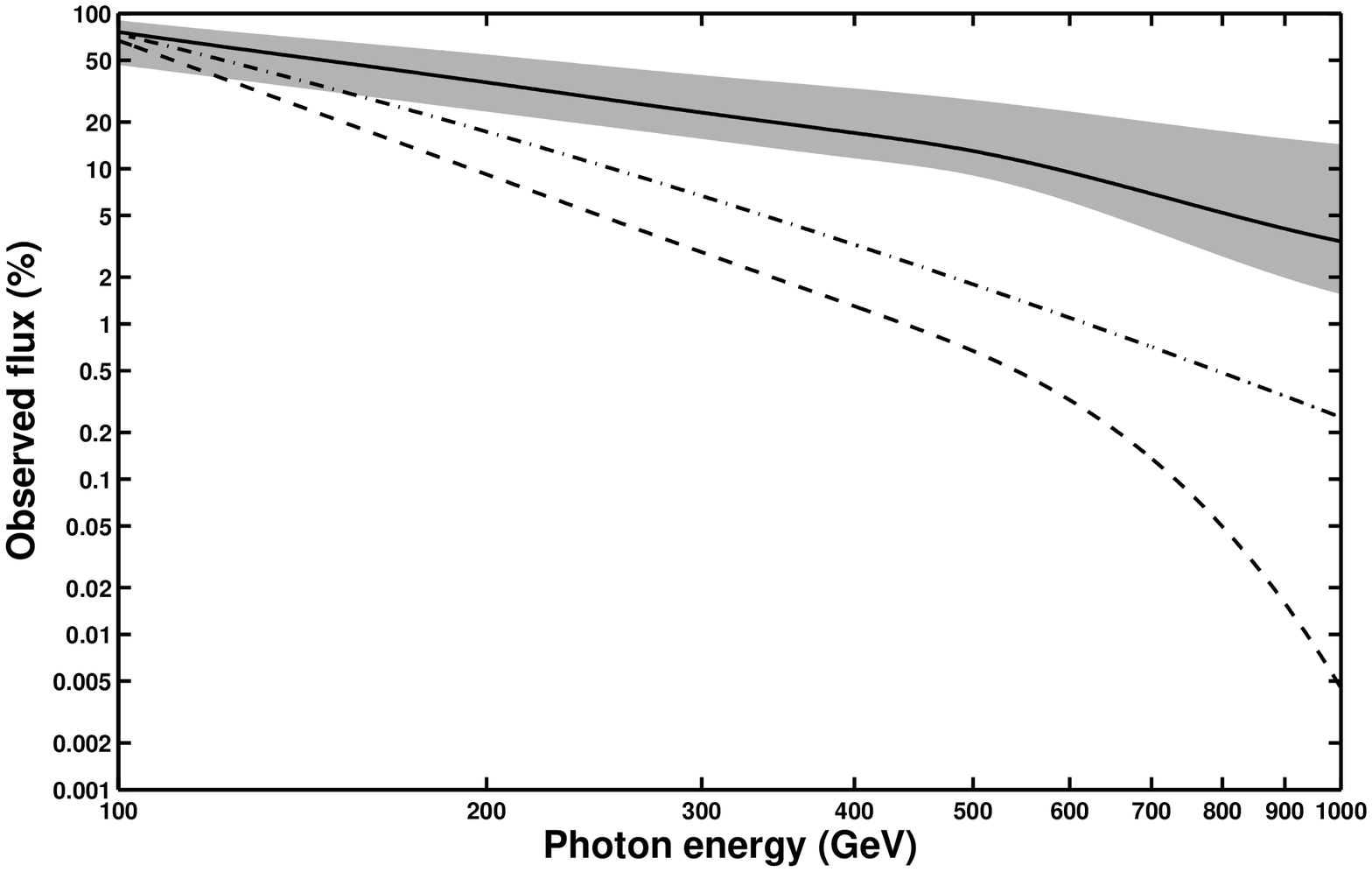}
\caption{\label{fig:comparison}
The two lowest lines give the fraction of photons surviving from a source at the same distance of 3C279 without the oscillation mechanism, for the ``best-fit model''~of~EBL (dashed line) and for the minimum EBL density compatible with cosmology~\cite{kneiske}. The solid line represents the 
prediction of the oscillation mechanism for \mbox{$B \simeq 10^{-9}\,{\rm G}$} and \mbox{$L_{\rm dom} \simeq 1\,{\rm Mpc}$} within the ``best-fit model'' of EBL. The gray band is the envelope of the results obtained by independently changing ${\bf B}$ and $L_{\rm dom}$ within a factor of 10 about their preferred values.}
\end{figure}

It is straightforward to solve the beam propagation equation once photon absorption by the EBL is taken into account and produces a finite photon mean free path
${\lambda}_{\gamma}$~\cite{ckpt}. In the strong-mixing regime, the probability for a photon to
become a $X$ boson after a distance $y$ reads
\begin{equation}
\label{hm2} 
P_{\gamma \to a}^{(0)}(y) \simeq \frac{1}{2} \, e^{- y/(2 {\lambda}_{\gamma})} \, {\rm sin}^2 \left( \frac{\delta y}{2 {\lambda}_{\gamma}} \right)~,
\end{equation}
whereas the probability that a photon remains a photon reads
\begin{equation}
\label{hm3} 
P_{\gamma \to \gamma}^{(0)}(y) \simeq \frac{1}{2} \, e^{- y/{\lambda}_{\gamma}} \, \left[ 1 
+ {\rm cos}^2 \left( \frac{\delta y}{2 {\lambda}_{\gamma}} \right) \right]~,
\end{equation}
where we have introduced the dimensionless parameter
\begin{equation}
\label{as1}
\delta \equiv \frac{B \, {\lambda}_{\gamma} }{M} \simeq 0.11 \left( \frac{B}{10^{-9}\, {\rm G}} \right) \left( \frac{10^{11} \, {\rm GeV}}{M} \right)
\left( \frac{{\lambda}_{\gamma}}{{\rm Mpc}} \right) .
\end{equation}
The Fazio-Stecker relation between the photon energy and the redshift of its horizon~\cite{figuraOriz} yields ${\lambda}_{\gamma} \simeq 450$~Mpc at $E = 500 \, {\rm GeV}$, so that $\delta \simeq 12.4$ for this case study.

Over distances \mbox{$y \gg L_{\rm dom}$}, the transition probabilities $P_{\gamma \to a}(y)$ and $P_{\gamma \to \gamma}(y)$ arise as the incoherent average of Eqs.~(\ref{hm2}) and (\ref{hm3}) over \mbox{$N \simeq (y/ L_{\rm dom})$} domains crossed by the beam, respectively.
Assuming (as before) that the beam propagates along the $y$ direction and choosing the $x$ and 
$z$ directions arbitrarily in the orthogonal plane, the problem becomes truly three-dimensional, because of the random orientation of the magnetic field.
Consequently, the beam state is described by the vector $({\gamma}_x, {\gamma}_z, a)$.

We have written down the propagation equations describing the absorption of photons due to the interaction with the EBL and their oscillations into $X$ bosons (and vice-versa).
Similarly to Ref.~\cite{ckpt}, we are led to the transfer equation
\begin{equation}
\left(\!
\begin{array}{c} 
\gamma_x \\ \gamma_z \\ a
\end{array} 
\!\!\right)
=
{\rm e}^{i E y}
\left[ \, T_0 \, {\rm e}^{\lambda_0 y}
+T_1 \, {\rm e}^{\lambda_1 y} + T_2 \, {\rm e}^{\lambda_2 y} \, 
\right]\!\!
\left(\!
\begin{array}{c}
\gamma_x \\ \gamma_z \\ a                      \label{eq:evolution}
\end{array}
\!\!\right)_{\!\!\!0}   \!
\end{equation}
%
% 
% \phantom{where}
% 
% \noindent
where
\begin{equation}
\begin{aligned}
\lambda_0 \equiv -\,\frac{1}{2\,{\lambda}_{\gamma}} \,,& \qquad
\lambda_1 \equiv - \, \frac{1}{4\,{\lambda}_{\gamma}} \, \left[ 1 +
\sqrt{1-4\,\delta^2} \right] \,\,,
\\
\lambda_2 \equiv& - \, \frac{1}{4\,{\lambda}_{\gamma}} \, \left[ 1 -
\sqrt{1-4\,\delta^2} \right] \,\,,
\end{aligned}
\end{equation}

\begin{widetext}
\begin{eqnarray}
&
T_0 \equiv \left( \begin{array}{ccc}
{\rm sin}^2 \theta & -\, {\rm cos} \theta \, {\rm sin} \theta & 0 \\
-\, {\rm cos} \theta \, {\rm sin} \theta & {\rm cos}^2 \theta & 0 \\
0 & 0 & 0
\end{array} \right) \,\,, \qquad
% \nonumber
% \\
T_1 \equiv \left( \begin{array}{ccc}
\frac{1+\sqrt{1-4\,\delta^2}}{2\,\sqrt{1-4\,\delta^2}}
\, {\rm cos}^2 \theta &
\frac{1+\sqrt{1-4\,\delta^2}}{2\,\sqrt{1-4\,\delta^2}}
\, {\rm cos} \theta \, {\rm sin} \theta &
-\,\frac{\delta}{\sqrt{1-4\,\delta^2}} \,{\rm cos} \theta \\
 \frac{1+\sqrt{1-4\,\delta^2}}{2\,\sqrt{1-4\,\delta^2}}
 \, {\rm cos} \theta \, {\rm sin} \theta &
\frac{1+\sqrt{1-4\,\delta^2}}{2\,\sqrt{1-4\,\delta^2}} \, {\rm sin}^2 \theta &
-\,\frac{\delta}{\sqrt{1-4\,\delta^2}} \,{\rm sin} \theta \\
\frac{\delta}{\sqrt{1-4\,\delta^2}} \,{\rm cos} \theta &
\frac{\delta}{\sqrt{1-4\,\delta^2}} \,{\rm sin} \theta &
-\,\frac{1-\sqrt{1-4\,\delta^2}}{2\,\sqrt{1-4\,\delta^2}}
\end{array} \right) \,\,, \nonumber
&
\\
&
T_2 \equiv \left( \begin{array}{ccc}
-\,\frac{1-\sqrt{1-4\,\delta^2}}{2\,\sqrt{1-4\,\delta^2}} \, {\rm cos}^2 \theta &
-\,\frac{1-\sqrt{1-4\,\delta^2}}{2\,\sqrt{1-4\,\delta^2}}
\, {\rm cos} \theta \, {\rm sin} \theta &
\frac{\delta}{\sqrt{1-4\,\delta^2}} \,{\rm cos} \theta \\
-\,\frac{1-\sqrt{1-4\,\delta^2}}{2\,\sqrt{1-4\,\delta^2}}
\, {\rm cos} \theta \, {\rm sin} \theta &
-\,\frac{1-\sqrt{1-4\,\delta^2}}{2\,\sqrt{1-4\,\delta^2}}
\, {\rm sin}^2 \theta &
\frac{\delta}{\sqrt{1-4\,\delta^2}} \,{\rm sin} \theta \\
-\,\frac{\delta}{\sqrt{1-4\,\delta^2}} \,{\rm cos} \theta &
-\,\frac{\delta}{\sqrt{1-4\,\delta^2}} \,{\rm sin} \theta &
\frac{1+\sqrt{1-4\,\delta^2}}{2\,\sqrt{1-4\,\delta^2}}
\end{array} \right) ~,
&
\end{eqnarray}
\end{widetext}
and $\theta$ is the angle between the $x$~axis and the extragalactic ${\bf B}$ in a single domain.
Starting with an unpolarized photon beam, we propagate it by iterating Eq.~(\ref{eq:evolution}) as many times as the number of domains crossed by the beam, taking each time a random value for the angle $\theta$ (this reflects the random orientation of ${\bf B}$), for $E = 500 \, {\rm GeV}$ and $z = 0.538$, corresponding to blazar 3C279. We next repeat such a procedure $10^.000$ times. Upon averaging over all these realizations of the propagation process, we find that about 13\% of the photons arrive to the Earth, resulting in an enhancement by a factor of about 20 with respect to the flux expected in the absence of the proposed oscillation mechanism; the comparison is made with the ``best-fit model'' described in in the first paper in
Ref.~[\citealt{kneiske}(a)].
The same calculation gives a fraction of 76\% at 100 GeV (to be compared to 67\% without the oscillation mechanism) and a fraction of 3.4\% at 1 TeV (to be compared to 0.0045\% without the oscillation mechanism) (cfr.~Fig.~\ref{fig:comparison}). In addition, we have checked the stability of our result against independent variations of ${\bf B}$ and $L_{\rm dom}$ within a factor of 10 about their preferred values. The resulting spectrum is represented by the gray band in Fig. 1. We remark that the standard deviation of the above averaging procedure lies well inside the gray band. Our predictions can be tested in the near future with IACTs.\vspace*{.1mm}

We thank Tanja Kneiske and Massimo Persic for suggestions and comments.


\begin{thebibliography}{99}
\bibitem{diffic}
The alternative possibility---concerning a modification of the source emission spectrum---turns out to be hardly compatible with current AGN models~\cite{fossati}.
This is due to the fact that even a small change in the photon mean-free-path ${\lambda}_{\gamma}$ entails a big change in the observed spectrum, because of the exponential dependence on ${\lambda}_{\gamma}$.
\bibitem{noaxion}
Notice that the axion is thereby excluded.
\end{thebibliography}
\end{document}